\documentstyle[11pt,aaspp4]{article}
\def\gsim{\;\rlap{\lower 2.5pt
 \hbox{$\sim$}}\raise 1.5pt\hbox{$>$}\;}
\def\lsim{\;\rlap{\lower 2.5pt
   \hbox{$\sim$}}\raise 1.5pt\hbox{$<$}\;}
\newcommand\beq{\begin{equation}}
\newcommand\eeq{\end{equation}}

\begin{document}
\title{Probing the Mass Fraction of MACHOs in Extragalactic Halos}
\author{Rosalba Perna \& Abraham Loeb}
\medskip
\affil{Harvard-Smithsonian Center for Astrophysics, 60 Garden Street,
Cambridge, MA 02138}

\begin{abstract}
Current microlensing searches calibrate the mass fraction of the Milky Way
halo which is in the form of Massive Compact Halo Objects (MACHOs). We show
that surveys like the Sloan Digital Sky Survey (SDSS) can probe the same
quantity in halos of distant galaxies. Microlensing of background quasars
by MACHOs in intervening galaxies would distort the equivalent width
distribution of the quasar emission lines by an amplitude that depends on
the projected quasar-galaxy separation. For a statistical sample of $\sim
10^5$ quasars as expected in the SDSS, this distortion is
detectable at the $\ga 2\sigma$ level out to a quasar-galaxy impact
parameter of several tens of kpc, as long as extragalactic halos are made
of MACHOs.  Detection of this signal would test whether the MACHO fraction
inferred for the Milky-Way halo is typical of other galaxies.

\end{abstract}


\centerline{submitted to {\it The Astrophysical Journal Letters}, 1997}

\section{Introduction}

One of the basic unknowns in cosmology is the fraction of the dark matter
mass which is in the form of compact objects, such as stellar remnants.
Paczy\'nski (1986) proposed monitoring of background stars in the Large
Magellanic Cloud (LMC) and searching for rare amplification events due to
microlensing as means of constraining the mass fraction of compact objects
in the Milky Way halo out to 50 kpc.  Current microlensing surveys suggest
that $\sim 50^{+30}_{-20}\%$ of the mass of the halo of the Milky Way
galaxy might be in the form of Massive Compact Halo Objects (MACHOs) with a
mass $\sim 0.5^{+0.3}_{-0.2} M_\odot$ (Alcock et al. 1996).  It remains to
be seen whether future studies confirm these early reports.

The search for brightening of {\it individual} background stars due to
microlensing by foreground MACHOs cannot be trivially extended beyond the
scale of the Milky Way halo. Images of galaxies at large distances include
many faint stars within each resolution element (pixel), most of which are
unlensed.  Alternatively, one can search for a weak temporal fluctuation
due to the chance amplification of one of the stars within a given pixel
(Gould 1996; Crotts 1996).  This method is
currently being applied to the nearest massive galaxy, M31 (Crotts \&
Tomaney 1996).

At yet larger distances, MACHOs can be detected through their effect on
quasars. The lensing cross-section of a MACHO increases in proportion to
its distance from the observer. Thus, despite the fact that quasars are
rare, the likelihood of a quasar which happens to be located behind an
extragalactic halo to be microlensed is $\sim 5$-$6$ orders of magnitude
larger than that of an LMC star behind the Galactic halo. This makes
quasars useful sources for the study of extragalactic halos as LMC stars
are for the Galactic halo. The only question is how can one identify the
existence of microlensing in the former case.

The crossing time of the lensing zone of solar mass MACHOs in the Galactic
halo is of the order of a month, and so microlensing of LMC stars can be
conveniently identified through the achromatic variability signal of an
otherwise steady stellar source.  However, if the same halo objects are
placed at cosmological distances, the duration of their lensing event is
stretched by $\sim 2$-$3$ orders of magnitude up to the inconvenient scale
of decades.  Moreover, the lensing events of quasars would be contaminated
by noise from the uncertain level of intrinsic variability of the quasars
on the same timescale (Maoz 1996).  Therefore, variability is not an
efficient search technique for microlensing by solar mass MACHOs in distant
halos.\footnote{Although Jupiter mass lenses could cause quasar
variability on the more convenient timescale of $\la$year (Hawkins \&
Taylor 1997), current microlensing searches (Alcock et al. 1996, Ansari et
al. 1996)
rule out the possibility that such objects contribute significantly to the
mass of the Galactic halo.}

An alternative method that does not require monitoring over time is to
search for a change in the equivalent width of the broad emission
lines of the quasar.  The characteristic Einstein radius of a solar
mass lens at a cosmological distance is $\sim 5\times 10^{16}~{\rm
cm}$, comfortably in between the scales of the continuum--emitting
accretion disk ($\la 10^{15}~{\rm cm}$) and the broad line region
($\sim 3\times 10^{17}~{\rm cm}$) of a bright quasar.  This implies
that a microlensing event would significantly amplify the continuum
but not the broad lines emitted by the quasar.  As a result, the
equivalent width distribution of the broad emission lines (Francis
1992) will be systematically distorted in a sample of microlensed
quasars (Canizares 1982, 1984).  Dalcanton et al.  (1994) used the
lack of redshift evolution in the equivalent width distribution of
quasars to limit the mean density parameter of MACHOs with a mass
$10^{-3}$--$60~M_\odot$ in the Universe, $\Omega_{_{\rm MACHOs}}\la
0.2$. This limit, however, does not exclude the possibility that
MACHOs account for most of the dark matter in galactic halos (or even
globally, if the Universe is open).  Since luminous stars cluster in
galaxies, it is possible that their dim counterparts are also
concentrated around galaxies.

The probability for microlensing depends on the projected separation
between the background quasar and the center of the intervening galaxy. In
this paper we propose to look for MACHOs in extragalactic halos, by
searching for the equivalent width distortion of quasar emission lines as a
function of the projected quasar-galaxy separation.  The forthcoming Sloan
Digital Sky Survey (SDSS; see http://www.astro.princeton.edu/BBOOK), will
catalog $\sim 10^5$ quasars and $\sim 10^6$ galaxies, and would provide an
ideal data base for such a study.

The outline of the paper is as follows. In \S 2 we describe our model
assumptions. In \S 3 we derive the expected distortion of the equivalent
width distribution of background quasars as a function of the MACHO mass
fraction in the intervening galactic halos.  In \S 4 we quantify the
expected lensing signal in SDSS, using Monte-Carlo realizations of the
quasar and galaxy fields. Finally, \S 5 summarizes our main conclusions.

\section{Model Assumptions}

We model a galactic halo as a singular isothermal sphere (SIS), whose
average surface mass density at a distance $\xi$ from the center is
$\Sigma(\xi)=\sigma^2/ (2G\xi)$. The one-dimensional velocity dispersion,
$\sigma$, is related to the luminosity, $L$, of the galaxy through the
Tully-Fisher (1977) relation, $\sigma/\sigma_\star=(L/L_\star)^\alpha$, with
$\alpha\approx 0.4$ in the $R$ band
(Strauss \& Willick 1995), $\sigma_\star\approx 170\;{\rm km}\;{\rm s}^{-1}$
and $L_\star=10^{10}\;L_\odot h^{-2}$.  It is convenient to normalize all
length scales in the lens plane by
\begin{equation}
\xi_0=4\pi\left(\frac{\sigma}{c}\right)^2\frac{D_{\rm ol}
D_{\rm ls}}{D_{\rm os}},
\label{eq:xi_0}
\end{equation}
where $D_{\rm ol}$, $D_{\rm os}$ and $D_{\rm ls}$ are the observer-lens,
the observer-source, and the lens-source angular diameter distances,
respectively.  These distances depend on the cosmological parameters; in
the following we assume $\Omega =1,\;\Lambda=0$ and $h=0.5$.  Defining
$x=\xi/\xi_0$ as the normalized image position in the lens plane, the
amplification factor of a point source by a SIS is
\begin{equation}
\langle\mu\rangle=\frac{x}{|x-1|}.
\label{eq:mu}
\end{equation}
Multiple images occur when $x<2$. For $x\geq2$, the single image position
is at $x=y+1$, where $y=\eta D_{\rm ol}/(D_{\rm os}\xi_0)$, and $\eta$ is the
distance of the source from the optical axis in the source plane (Schneider
et al. 1992).

Equation~(\ref{eq:mu}) yields the amplification factor $\mu$ only in the
case where the mass distribution is perfectly smooth. However, in the
presence of MACHOs one obtains a probability distribution of amplification
factors at any given value of $x$.  The average amplification with this
distribution must still be given by equation~(\ref{eq:mu}) to guarantee
flux conservation.  Since the lensing zone of a solar mass MACHO is six
orders of magnitude smaller than the length scale over which the global
$\Sigma$ varies, the MACHOs can be treated locally as a field of randomly
distributed point masses with some average surface density $\Sigma$.  We
then define $p(\mu,x,\epsilon)d\mu$ as the probability that a point source
whose line-of-sight passes at an impact parameter $x$ relative to the
center of a galaxy is amplified by a factor between $\mu$ and $\mu+d\mu$.
This probability depends on the MACHO mass fraction $\epsilon$, which we
take as a constant. Schneider (1987) and Bartelmann
\& Schneider (1990) derived the following analytical expression which
agrees reasonably well with results from numerical simulations,
\begin{equation}
p(\mu,x,\epsilon)=\left[\frac{C\times (\mu-\mu_*)^\eta}{\mu^{\eta+3}}
+ A\times (\mu-\mu_*)\exp(-B\mu)d\mu\right] H(\mu-\mu_*)d\mu.
\label{eq:pmu}
\end{equation}
Here $H(...)$ denotes the Heavyside step function; $A$, $B$, and $C$ are
functions of $x$ --the first two are determined by the constraints of
normalization and flux conservation and the third by the limit of high
amplification (Schneider 1987).  The parameter $\eta$ is chosen so as to
make $A$ and $B$ well-defined and positive for all possible values of $x$.
The value of $\eta$ rises monotonically with decreasing $\epsilon$, because
the probability distribution approaches a $\delta$--function as the MACHO
fraction goes to zero.  The parameter $\mu_\star$ is a function of $\epsilon$
such that it approaches $\langle\mu\rangle$ in the limit of
$\epsilon\rightarrow 0$ (see Bartelmann \& Schneider 1990 for full details).

Our goal is to quantify the distortion of the equivalent width distribution
of quasar emission lines due to microlensing by MACHOs in halos of
foreground galaxies.  For a given magnitude-limited sample of galaxies and
quasars, we need to select the galaxy and quasar properties at random.  We
assume that the local luminosity function of galaxies has the Schechter
form,
\begin{equation}
\phi_G(L)dL=\phi_\star\left(\frac{L}{L_\star}\right)^\nu\exp\left\{
-{L\over L_\star}\right\}
d\left(\frac{L}{L_*}\right),
\label{eq:lum}
\end{equation}
with $\phi_\star=0.018 \;(h/0.5)^3 \;{\rm Mpc}^{-3}$ and $\nu=0.3$ in the
$R$ band (Lin et al.  1996). Since a typical survey contains galaxies at
very low redshifts (e.g. $\langle z\rangle \sim 0.1$ for the SDSS), we
ignore evolutionary corrections and $k$-corrections to this luminosity
function and scale the density of galaxies in each luminosity bin by
$(1+z)^3$.  On the other hand, the quasar density peaks at high redshifts
$z=2$--4 (see Fig. 2 in Shaver et al.  1996). Because of the disparity
between the lens and source redshifts, we can assume that all the quasars
are at an effective redshift $z_{\rm s}\approx 3$ and make use of their
well-determined total number count.  The inaccuracy introduced by this
approximation to the lensing calculation is small ($\sim 10$--$20\%$) and
not significant relative to the observational uncertainties in the redshift
evolution of the quasar luminosity function. In this approach, we only need
to use the integral number counts of quasars as a function of $B$ band
flux, which is well described by a broken power-law
\begin{equation}
  N(>S) = C'\times \left\{ \begin{array}{ll}
\left({S}/{S_0}\right)^{-\beta_1} & \hbox{if $S\le S_0$} \\
\left({S}/{S_0}\right)^{-\beta_2} & \hbox{if $S>S_0$} \\
\end{array}\right.\;,
\label{eq:19}
\end{equation} 
where $\beta_1\approx 1$ and $\beta_2\approx 2.5$ (e.g., Setti, 1984;
Marshal, 1985), and $C'\approx 1/{\rm deg}^{2}$ (Hartwick \& Schade 1990).
The break flux $S_0$ corresponds to a $B$ magnitude $\sim 19.5$. We can
safely ignore any correlation between the quasars and the galaxies because
of the clear separation between their redshifts.

\section{Signature of Microlensing}
  
The flux of a quasar can be significantly amplified by microlensing only
if the size of its emission region is smaller than the projected Einstein
radius of the lens in the source plane, $r_{\rm E}$.  The maximum
amplification of a circular source of radius $r_s$ and uniform brightness
is given by [Schneider et al.  (1992), p. 38]
\begin{equation}
\mu_{\rm max}=\sqrt{1+4(r_{\rm E}/r_s)^2}\; .\label{eq:mumax}
\end{equation}  
At cosmological distances, the Einstein radius of a star of mass $M_{\rm
star}$ obtains the characteristic value $r_{\rm E}\sim 5\times
10^{16}~(M_{\rm star}/M_\odot)^{1/2}~{\rm cm}$.  In comparison, the optical
continuum emission of quasars is believed to originate from a compact
accretion disk.  The UV bump observed in quasar spectra is often
interpreted as thermal emission from an accretion disk with a surface
temperature $T_{\rm disk}\equiv 10^5 T_5~{\rm K}$, where $T_5\sim 1$ (e.g.
Laor 1990), and so the scale of the disk emission region must be $\sim
10^{15}~{\rm cm}\times T_5^{-2} L_{46}^{1/2}$, where $L_{46}$ is the
corresponding luminosity of the quasar in units of $10^{46}~{\rm
erg~s^{-1}}$.  Thus, for lens masses $M_{\rm star}\gg 10^{-3}M_\odot$, the
continuum source is much smaller than the Einstein radius and could
therefore be amplified considerably.  This expectation is indeed confirmed
in the nearby lens of Q2237+0305, where variability due to microlensing has
been observed (Wambsganss et al. 1990; Rauch \& Blandford 1991; Racine
1991; see also Gould \& Miralda-Escud\'e 1996).  On the other hand,
reverberation studies of the time lag between variations in the continuum
and the line emission in active galactic nuclei indicate that the broad
emission lines of quasars originate at a distance of $\sim 3\times
10^{17}~{\rm cm}~L_{46}^{1/2}$ (Peterson 1993; Maoz 1996; Netzer \&
Peterson 1997).  For a solar mass lens and a quasar with $L_{46}\sim 1$,
equation~(\ref{eq:mumax}) implies that the maximum line amplification
differs from unity by only $\la 10\%$.  This argues that microlensing of
the broad line region by a single star can be neglected.  The broad line
region would, however, be macrolensed by a factor $\langle\mu\rangle$ due
to the average effect of the galaxy as a whole.  As a result of this
differential amplification of lines and continuum, the equivalent width of
the lines will change during a microlensing event.

More specifically, let us define ${I}_{\nu}(\lambda_0)$ to be the intensity
of the quasar continuum in the neighborhood of the wavelength $\lambda_0$
of a particular emission line and $\Delta{I}_{\nu} (\lambda)$ to be the
difference between the total (line+continuum) intensity and the continuum.
If the continuum is amplified by a factor $\mu$, then its observed
intensity changes to $\mu{I}_{\nu}(\lambda_0)$. The intensity of the lines
is enhanced by a factor $\langle\mu\rangle$, given by equation
(\ref{eq:mu}).  Consequently, the equivalent width (EW) of the emission
line,
\begin{equation}
W\equiv\int\frac{\Delta{I}_{\nu}}{{I}_{\nu}(\lambda_0)}
d\lambda
\label{eq:wl}
\end{equation}
is changed by a factor $\langle\mu\rangle/\mu$, namely
\begin{equation}
W=W_0\frac{\langle\mu\rangle}{\mu}
\label{eq:wnew}
\end{equation}
where $W_0$ is the intrinsic EW of the unlensed
quasar.

Even in the absence of lensing, quasars do not possess a single EW value in
their emission lines but rather show a wide probability distribution of EW
values (Francis 1992), which we define as $P(W_0)$ and model by a
log-Gaussian form,
\begin{equation}
P(W_0)=\frac{1}{\sqrt{2\pi\sigma_{W}^2}W_0} \exp\left\{
{-[\ln(W_0)-\omega]^2\over {2\sigma_{W}^2}}\right\}.
\label{eq:pw0}
\end{equation}
The parameters $\omega$ and $\sigma_{W}$ obtain different values for
different emission lines.  The distribution in equation~(\ref{eq:pw0}) is
assumed to be observed in the absence of lensing. However, if the line of
sight to a quasar passes at an impact parameter $x$ from the center of a
galaxy with a MACHO fraction $\epsilon$, then its equivalent width is drawn
from the distribution
\begin{equation}
P(W,x,\epsilon)=\int_1^{\infty}d\mu\; p(\mu,x,\epsilon)
P\left(\frac{\mu}{\langle\mu\rangle} W\right)
\frac{\mu}{\langle\mu\rangle}.
\label{eq:pwl}
\end{equation}

Figure 1 shows the probability distribution in equation~(\ref{eq:pwl}) as a
function of $x$, for $\epsilon=0.3$ (panel a), $\epsilon=0.5$ (panel b) and
$\epsilon=1$ (panel c).  The unlensed distribution was selected for the
MgII line, based on the data in Francis (1992). The microlensing signal is
similar for other lines (Perna \& Loeb 1997).  Naturally, the systematic
distortion of the lensed distribution increases as $x$ decreases, because
of the corresponding increase in the surface density of MACHOs.  However,
the number of observed quasars declines as $x$ becomes smaller, and so the
signal suffers from an increasing statistical noise in this limit.  In
order to properly evaluate the signal-to-noise ratio we need to combine the
above effects.

Let us consider a galaxy with luminosity $L$ at redshift $z$ and a
population of background quasars with a flux $>S_{\rm QSO}$ and an average
number per solid angle $N(>S_{\rm QSO})$.  We define $\theta_s$ to be the
angle between the line-of-sight to a given quasar and the optical axis of
the galaxy.  The number of quasars within a ring of width $d\theta_s$
around $\theta_s$ is $N(>S_{\rm QSO})2\pi\theta_s d\theta_s$.  Using
$\theta_s=\eta/D_{\rm os}$, and defining $\theta_0\equiv
\xi_0/D_{\rm ol}$, this number can be expressed as
\begin{equation}
N(>S_{\rm QSO})2\pi\frac{\eta d\eta}{D_{\rm os}^2}= N(>S_{\rm
QSO})2\pi\frac{\xi_0^2}{D_{\rm ol}^2}ydy= N(>S_{\rm QSO},
x,\epsilon) 2\pi\theta^2_0 (x-1)dx.
\label{eq:1}
\end{equation}
The last step used the relation between $x$ and $y$ for a single image in
the SIS model and incorporated amplification bias through the dependence of
the local density of quasars with observed flux $>S_{\rm QSO}$ on $x$,
\begin{equation}
N(>S_{\rm QSO},x,\epsilon)=\frac{1}{\langle\mu(x)\rangle}
\int_1^{\infty}d\mu\;p(\mu,x,\epsilon)
N(>\frac{S_{\rm QSO}}{\mu}).
\label{eq:2}
\end{equation}
Equation~(\ref{eq:1}) gives the number of quasars per unit solid angle
inside a ring of radius $x$ and width $dx$ around the center of a single
galaxy.  To obtain the total number of quasars per solid angle which are
seen within a distance $dx$ of $x$ of all galaxies up to a limiting flux
$S_{\rm Gal}$, one has to integrate over the galaxy distribution, both in
luminosity and in redshift,
\begin{eqnarray}
N(>S_{\rm QSO},>S_{\rm Gal},x,\epsilon)dx &=&\frac{2\pi
(x-1)dx}{\langle\mu(x)\rangle}
\int_1^{\infty}d\mu\;p(\mu,x,\epsilon)N\left(>\frac{S_{\rm QSO}}{\mu}\right)
\nonumber\\
&\times&\int_0^{z_{\rm s}}dz\left|c\frac{dt}{dz}\right|(1+z)^3
\int_{L_{\rm Gal}}^\infty dL\;\phi_G(L)\xi_0^2(L,z).
\label{eq:numx}
\end{eqnarray}
Here we have implicitely assumed that the rings around different galaxies
do not overlap. As shown later, the signal-to-noise ratio of the
microlensing signature peaks at small values of $x$, where this assumption
appears to be well satisfied.  If $S_{\rm Gal}$ is the minimum flux needed
to detect a galaxy, then $L_{\rm Gal}$ in equation~(\ref{eq:numx}) is given
by $L_{\rm Gal}=4\pi D^2_L(z)S_{\rm Gal}$, with the luminosity distance
$D_{\rm L}(z)=(1+z)^2 D_{\rm ol}(z)$.

\section{Applications and Results}

Next, we apply the results from \S 3 to simulated data for a survey similar
to SDSS.  Let $S_{\rm QSO}$ and $S_{\rm Gal}$ be the minimum fluxes needed
for the detection of quasars and galaxies, respectively, and let
$\Delta\Omega$ be the total solid angle surveyed on the sky. By definition,
the overall (sky averaged) source count of quasars is not influenced by the
amplification bias.  Hence, the total number of detected quasars is given
by $N_{\rm TOT} = N(>S_{\rm QSO}) \Delta\Omega$, with $N(>S)$ from
equation~(\ref{eq:19}).

To simulate the results of real observations, we start by generating
$N_{\rm TOT}$ random numbers drawn from the distribution
\begin{equation}
P(x) \equiv \frac{1}{N(>S_{\rm QSO})} N(>S_{\rm QSO},>S_{\rm
Gal},x,\epsilon)
\label{eq:px}
\end{equation}
such that
\begin{equation}
\int_{0}^{x_{\rm max}}dx\;N(>S_{\rm QSO},>S_{\rm Gal},x,\epsilon) = 
N(>S_{\rm QSO})
\label{eq:norm}
\end{equation}
with $N(>S_{\rm QSO},>S_{\rm Gal},x,\epsilon)$ given by
equation~(\ref{eq:numx}). In our actual calculation, the value of the lower
limit of integration has been set to 2 to avoid the complications arising
from multiple images. Since the fraction of quasars which are multiply
imaged is very small ($\la 1\%$), our statistical analysis is not affected
by this assumption. The value of $x_{\rm max}$ is fixed by equation
(\ref{eq:norm}) to reflect the crude boundary beyond which the halos of
foreground galaxies overlap; its precise value has little impact on our
conclusions since the signal-to-noise ratio is high only for values of $x\ll
x_{\rm max}$.

We have generated a source catalog at random based on equation
(\ref{eq:px}), with equivalent widths $W$ chosen according to the
probability distribution in equation~(\ref{eq:pwl}).  This data set is
chosen to represent a mock realization of SDSS, for which we evaluate
the signal-to-noise ratio of the microlensing signature assuming a
spectral resolution for the lines of $\sim 5$\AA.  Our analysis of
this catalog involves several steps.  First we bin all sources based
on their equivalent width, $W_i$, and derive the ``average''
probability histogram $P_{\rm ave}(W_i)$, which reflects
equation~(\ref{eq:pw0}) to within the statistical noise.  We then bin
the sources based on their lowest value of $x$ relative to a
foreground galaxy.  (Note that the association of a galaxy to a quasar
based on the minimum $x$-value does not necessarily pick the closest
galaxy to the quasar on the sky.)  The equivalent widths of the
quasars in each $x_i$-bin are then binned, and a $\chi^2$ analysis is
performed to test the significance of the deviation of the data in
each bin from the average distribution $P_{\rm ave}(W_i)$.  More
specifically, we define ${\chi}^2=\sum_i [{({\hat
n}_i-n_i)^2}/{n_i}]$, where ${\hat n}_i$ is the number of quasars in
the equivalent-width bin $i$ within a given $x$-bin and $n_i$ is the
expected number according to the average distribution $P_{\rm ave}$.
The end point of each $x_i$-bin is varied (starting from the end of
the previous $x_{i-1}$ bin) so as to minimize $P(\chi^2)$.  The
smaller $P(\chi^2)$, the more significant is the deviation of the data
in each bin from $P_{\rm ave}$.

Figures (2a), (b) and (c) show $P(\chi^2)$ as a function of the binned
$x$ for simulated sets of data with $z_{\rm s}=3$, and magnitude
detection thresholds of $m_{\rm Gal}\approx 22~R$-mag and $m_{\rm
QSO}\approx 22~B$-mag, giving a total quasar count of $N_{TOT}\sim
10^{5}$, as expected for the forthcoming SDSS.  In panel (a) the MACHO
fraction is $\epsilon=0.3$, in panel (b) $\epsilon=0.5$, and in (c)
$\epsilon=1$. The first case corresponds to the situation where the
halo dark matter is dominated by a smooth dark matter component (e.g.,
elementary particles), and the last case to a situation where all the
dark matter is in the form of MACHOs.  The difference between the
three panels is apparent. For a value of $\epsilon < 0.5$, $P(\chi^2)$
is of order of several tenths, implying that the microlensing signal
is not statistically significant.  However, for $\epsilon\sim 0.5$,
the value of $P(\chi^2)$ at small $x$ indicates that the EW
distribution starts to become statistically different from $P_{\rm
ave}$ as a result of microlensing. The difference between these
distributions is strong when the MACHO fraction is close to unity.
Figure (2c) shows that the equivalent widths of quasars whose lines of
sight intercepts a galaxy at an impact parameter $x\la 10$ from its
center, can be rejected as being drawn from the distribution $P_{\rm
ave}$ with a confidence level $\ga 2\sigma$ (the threshold level
indicated by the dashed horizontal line). We have found this result to
be robust using several realizations of the mock SDSS catalog.

The length scale associated with a given value of $x=\xi/\xi_0$ depends on
the luminosity $L$ and redshift $z_{\rm l}$ of the lensing galaxy, and 
on the source redshift $z_{\rm s}$; for $z_{\rm s}=3$,
\begin{equation}
\xi_0 =96 \frac{(L/L_\star)^{0.4}}{(1+z_{\rm l})}
\left[\frac{3}{2\sqrt{1+z_{\rm l}}}-\frac{1}{1+z_{\rm l}}
-\frac{1}{2}\right]\; {\rm kpc}.
\end{equation}
For example, an $L_\star$ galaxy obtains $\xi_0= 3.7$ kpc at $z_{l} =0.5$.
Therefore, detection of a microlensing signal at $x\sim 10$ would probe
galactic halos out to several tens of kpc, similar to the local
microlensing searches.  Since this scale extends well beyond the luminous
cores of galaxies, our proposed method would sample the dark matter content
of the intervening halos.

\section{Conclusions}

We have shown that the MACHO fraction in extragalactic halos can be
inferred from the distortion in the equivalent width distribution of
background quasars as a function of their separation from foreground
galaxies. This method could be used to test whether the
properties of the Milky Way halo as inferred from local microlensing
surveys are characteristic of other galaxies.

For the magnitude limit and survey area of SDSS, we find that if galactic
halos are made of MACHOs, the ratio between the microlensing signal and the
statistical rms noise is $\ga 2$ out to a halo radius of several tens of
kpc (see Fig. 2).  Larger surveys might be able to probe the signature of
halos with a smaller mass fraction of MACHOs.

\acknowledgements
We thank Matthias Bartelmann for providing us with a computer routine which
calculates the probability distribution of amplifications.  This work was
supported in part by the NASA ATP grant NAG5-3085 and the Harvard Milton
fund (for AL) and by a fellowship from the university of Salerno, Italy
(for RP).

\newpage
\begin{figure}[t]
\vspace{2.6cm}
\includegraphics{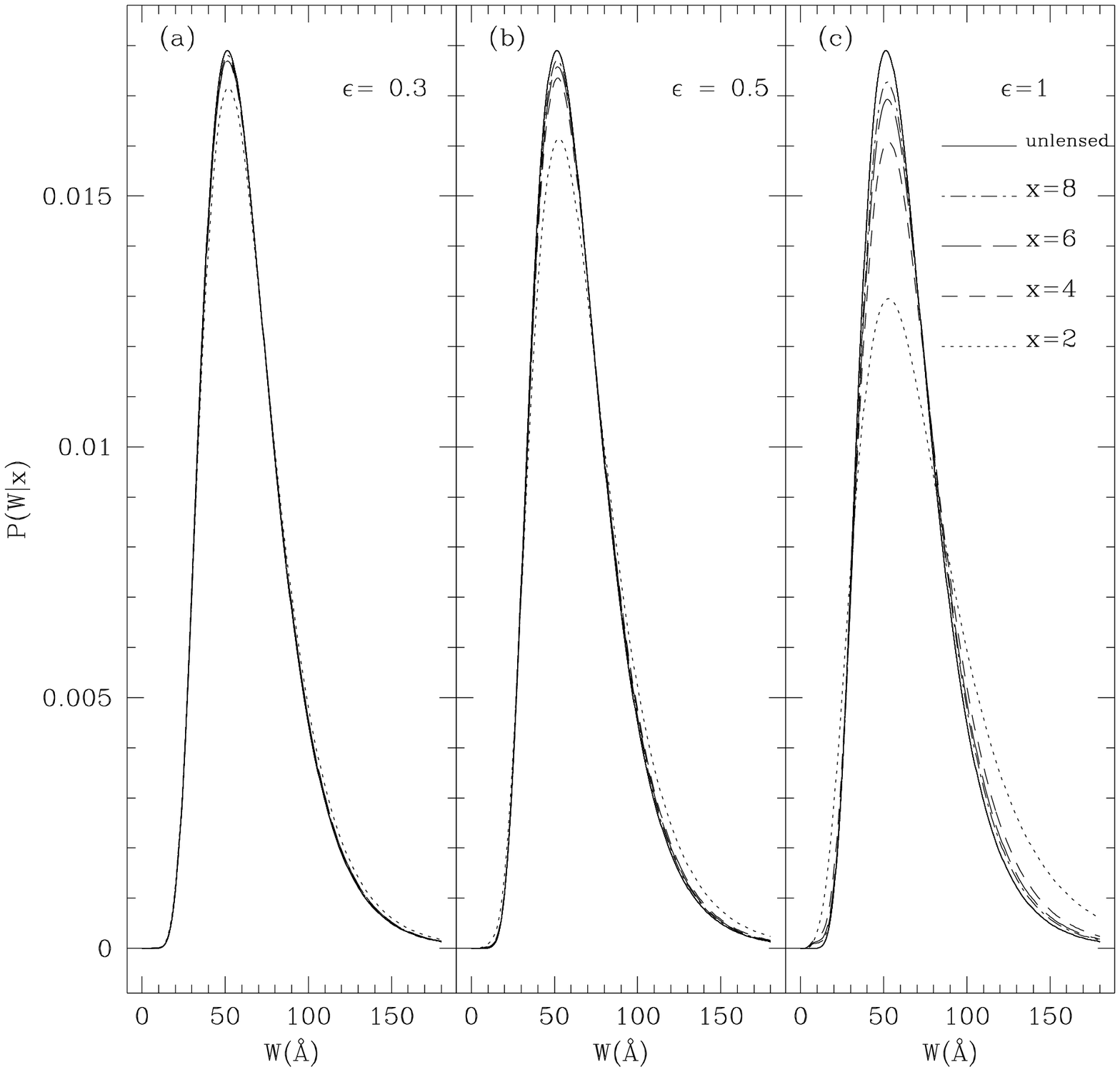}
\vspace*{4.5in}
\caption{The equivalent width distribution for the MgII emission
of quasars. The solid line reflects the assumed unlensed distribution.  The
other curves show the distributions expected for a sample of microlensed
quasars whose line of sight passes at an impact parameter $x=2$ (dotted
line), $x=4$ (dashed line), $x=6$ (long-dashed line), and $x=8$
(dotted-dashed line) from the center of a foreground galactic halo. In
panel (a) the MACHO fraction is $\epsilon=0.3$, in panel (b)
$\epsilon=0.5$, and in (c) $\epsilon=1$. }
\label{fig:1}
\end{figure}

\newpage
\begin{figure}[t]
\vspace{2.6cm}
\includegraphics{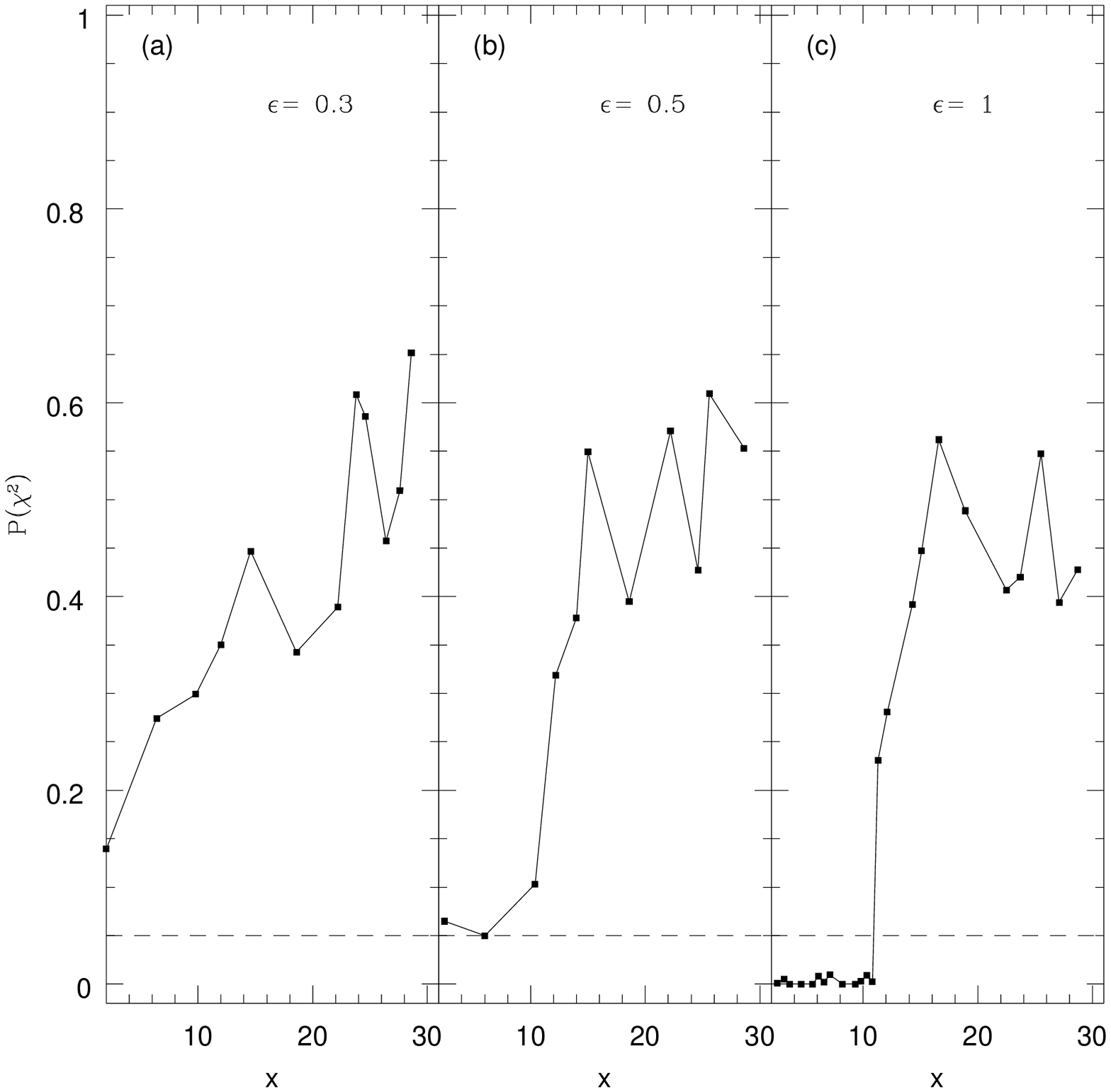}
\vspace*{4.5in}
\caption{The $\chi^2$-probability as a function of the binned $x$
for a simulated SDSS catalog in which the equivalent widths and the
quasar-galaxy separation $x$ of $\sim 10^5$ quasars are measured.  The
value of $P(\chi^2)$ indicates the probability that the EW data in each
$x$-bin are described by the average distribution $P_{\rm ave}(W_i)$,
obtained by considering all the EW's, with no reference to quasar-galaxy
separations. In panel (a) $\epsilon=0.3$, in panel (b) $\epsilon=0.5$, and
in (c) $\epsilon=1$. }
\label{fig:2}
\end{figure}


\begin{references}
\reference{}
Alcock, C., et al. 1996, astro-ph/9606165
\reference{}
Ansari, R. et al. 1996, A \& A, 314, 94
\reference{}
Bartelmann, M. \& Schneider, P. 1990, A\&A, 239, 113
\reference{}
Canizares, C. R. 1982, ApJ, 263, 508
\reference{}
Canizares, C. R. 1984, in Quasars and Gravitational Lenses 
(Liege: Univ. Liege, Cointe-Ougree), 126
\reference{}
Crotts, A. P. S., presented at the International Conference on
{\em Dark and Visible Matter in Galaxies}, Sesto Pusteria, Italy,
2--5 July 1996, astro-ph/9610067
\reference{}
Crotts, A. P. S., Tomaney, A. B. 1996, ApJ, 473L, 87
\reference{}
Dalcanton, J. J., Canizares, C. R., Granados, A. \& Stocke, J. T.
1994, ApJ, 568, 424
\reference{}
Francis, P. J. 1992, ApJ, 405, 119
\reference{}
Gould, A. 1996, ApJ, 470, 201
\reference{}
Gould, A., \& Miralda-Escud\'e, J. 1996, preprint astro-ph/9612144
\reference{}
Hartwick, F. D. A., \& Schade, D. 1990, ARA\&A, 28, 437
\reference{}
Hawkins, M. R. S. \& Taylor, A. N. 1997, ApJ, 482L, 5
\reference{}
Laor, A. 1990, MNRAS, 246, 369  
\reference{}
Lin, H., Kirshner, R. P., Schectman, S. A., Landi, S. D.,
Oemler, A., Tucker, D. L. \& Schecter, P. L. 1996, ApJ, 464, 60
\reference{}
Maoz, D. , to appear in the proceedings of IAU Colloquium 159,
Shangai, June 1996, preprint astro-ph/9609174 
\reference{}
Marshall, H. L. 1985, ApJ, 299, 109
\reference{}
Netzer, H., \& Peterson, B. 1997, preprint astro-ph/9706039
\reference{}
Paczy\'nski, B. 1986, ApJ, 304, 1
\reference{}
Perna, R \& Loeb, A. 1997, submitted to ApJ, preprint astro-ph/9701226
\reference{}
Peterson, B. M. 1993, PASP, 105, 247
\reference{}
Racine, R. 1991, AJ, 102, 454
\reference{}
Rauch, K. P., \& Blandford, R. D. 1991, ApJ, 381, L39
\reference{}
Shaver, P. A., Wall, J. V., Kellermann, K. I.,
Jackson, C. A. \& Hawkins, M. R. S. 1996, Nature, 384, 439
\reference{}
Schneider, P. 1987, A\&A, 179, 80
\reference{}
Schneider, P., Ehlers, J., \& Falco, E. E. 1992, Gravitational
Lenses (Heidelberg: Springer)
\reference{}
Setti, G. 1984, in: {\em X-ray and UV emission from Active Galactic
Nuclei}, ed. by W. Brinckmann and J. Tr\"umper, Garching
\reference{}
Strauss, M. A. \& Willick, J. A. 1995, {\em Physics Reports}, 261, 271
\reference{}
Tully, R. B. \& Fisher, J. R. 1977, A\&A, 54, 661 
\reference{}
Wambsganss, J., Paczy\'nski, B., \& Schneider, P. 1990, ApJ, 358, L33

\end{references}
\end{document}